\newcommand{\la}{\langle}
\newcommand{\ra}{\rangle}

\newcommand{\beq}{\begin{eqnarray}}
\newcommand{\eeq}{\end{eqnarray}}

\renewcommand{\theequation}{\thesection.\arabic{equation}}
\newcommand{\Sigpi}{\Sigma _{\pi N}}

\newcommand{\do}{^{\circ }}

\newcommand{\chis}{ChS\,\,}
\newcommand{\pipi}{$\pi$-$\pi$\, }

\newcommand\con{\langle \bar {q} q \rangle }

\newcommand{\btem}{\bibitem}
\newcommand{\TH}{T.\ Hatsuda}
\newcommand{\TK}{T.\ Kunihiro}
\newcommand{\MPL}{Phys.\ Lett.\ {\bf B}}
\newcommand{\MPTP}{Prog.\ Theor.\ Phys.}
\newcommand{\MPR}{Phys.\ Rev.}
\newcommand{\MPRL}{Phys.\ Rev. \ Lett.}

\newcommand{\HK}{T. Hatsuda and T. Kunihiro}
\newcommand{\KH}{T. Kunihiro and T. Hatsuda}
\documentstyle[seceq,epsf,twoside]{ptptex}
\notypesetlogo  
\title{Roles of Chiral Symmetry and the Sigma Meson in Hadron and Nuclear 
Physics}
\author{%
Teiji Kunihiro
}
\inst{%
Yukawa Institute for Theoretical Physics, Kyoto University,\\
Sakyoku, Kyoto 606-8502, Japan
}
\abst{%
We first review the recent accumulating evidences of the 
existence of a scalar-isoscalar meson with the mass
500 to 800 MeV which may be identified with 
the sigma meson as 
the quantum fluctuation of the amplitude of the 
chiral order parameter $\con $.
We indicate that phase shift analyses which respect 
chiral symmetry (\chis), 
analyticity and crossing symmetry of the scattering 
amplitude show the sigma meson 
pole in the $s$-channel 
as well as the $\rho$ meson pole in the $t$-channel 
in the \pipi scattering in the $I=J=0$ channel.
We emphasize  that the existence of the $\sigma$ resonance
does not contradict with the success of the 
chiral perturbation theory; phenomenological difficulties with 
the renormalizable linear sigma model do not necessarily deny 
the validity of the linear 
representation of \chis of QCD as given by the NJL-like models
which not only admit the $\sigma$ resonance but also 
reproduce the coupling
constants $L_i$ and $H_i$ appearing the nonlinear chiral lagrangian.
We give some examples of the hadronic 
phenomena which are naturally accounted for
with the $\sigma$ meson.
We show that the the $\sigma$ meson as the amplitude fluctuation 
of the chiral order parameter may be more clearly identified 
than in free space in hot and/or dense matter,  even in finite nuclei
where  partial restoration of \chis may be realized.
}

\begin{document}
\maketitle

\setcounter{tocdepth}{4}

\section{Introduction}
The low-energy  hadronic world is characterized by 
the  dynamical breaking of chiral symmetry (\chis), 
 $U_A(1)$ anomaly,
 explicit $SU_V (3)$ breaking,
 success of the constituent quark model,
 OZI rule and its  violation  in mesons and baryons,\,
 vector meson dominance, and so on,
even  apart from the   confinement of the colored quarks and gluons.
Although it is still a big challenge to explain
the above mentioned problems directly from QCD,
a semi-phenomenological but unified description of the above facts
may be  possible  using a chiral effective model
\cite{vogl,HK94,bijnens}. 
As was first discussed by Nambu\cite{NJL},
the collective nature of the vacuum and some hadrons 
related to the dynamical breaking of 
\chis is essential for giving the unified description, 
especially in realizing the chiral 
quark picture\cite{CHQM,ptp85}.
In Ref.2), the  dominant role of chiral symmetry in 
low-energy hadron dynamics was emphasized
\footnote{The $\Delta I= 1/2$ rule both 
in the meson and baryon decays may be a reflection of some 
collective nature of the dynamics originated from \chis;
 the $\sigma$  meson\cite{morozumi}
 and the diquarks\cite{diquark} may represent such collectiveness.}.
One may also recognize that \chis plays an essential role 
in making nuclei stable hence our very existence.
Notice that nuclei are only stable bound systems 
in the hadron world. One may say that this stability  is 
 due to the chiral symmetry of QCD; see \S 6 of Ref. 9).

The basic observation on which the  whole discussions in this 
report are based is that the dynamical breaking of \chis is 
a phase transition of the QCD vacuum with an order 
parameter $\con $ $\sim \sigma_0$,
hence there may exist  collective excitations corresponding to 
the quantum fluctuations of the order parameter:
The quantum fluctuation of the phase of the order parameter
is the pion, while 
the $\sigma$ meson as we call here is nothing but the quantum
fluctuation of the
amplitude of the chiral condensate. Therefore, exploring  the
existence of the $\sigma$ meson and its possible roles in the 
hadron world are of fundamental importance 
for understanding of the nonperturbative
structure of the QCD vacuum.

\section{Low-energy QCD and the $\sigma$ meson}

\subsection{The $\sigma$ meson as the quantum fluctuation of  chiral
 order parameter}

Theoretically, the scalar quark condensate 
$\con $=$\sigma_0 $ is determined as the value  
where the effective  potential (free energy) 
${\cal V}(\sigma)$ takes the minimum.
The $\sigma$ meson is the   particle representing the quantum fluctuation
$\tilde {\sigma}\sim \la :(\bar q q)^2:\ra$ as stated above;
$\sigma = \sigma _0 + \tilde{\sigma}$.
In this sense, the $\sigma$ meson 
is analogous to the Higgs particle in the standard
 model, 
where the Higgs field is the order parameter, and the quantum 
fluctuation of the field around the minimum point 
of the Higgs potential 
or the effective potential is the Higgs particle 
in the present world.
\footnote{The Nambu-Goldstone (NG) bosons in the standard model are 
absorbed away to the longitudinal component of the 
gauge bosons, while the NG boson in QCD is the pion.}

\subsection{Chiral perturbation theory and the $\sigma$ meson}

Some effective theories\cite{HK94} including the ladder
QCD\cite{scadron} 
predict the $\sigma$ meson mass $m_{\sigma}= 500-800$ MeV.
Furthermore,  Weinberg's mended symmetry\cite{weinberg} also
leads to the existence of the $\sigma$ meson 
and the degeneracy 
 of it with the $\rho$ meson.

The Nambu-Jona-Lasinio(NJL)-like  models
\cite{NJL,vogl,klevansky,HK94,bijnens}
are known to work well 
 as an effective theory
which well describe the chiral properties of of the low-energy 
hadronic world including resonance phenomena and the processes
incorporating chiral anomaly. In these models, 
 \chis is realized linearly 
and can incorporate the vector mesons as well.
It is worth emphasizing that the NJL-like models
not only  predict the $\sigma$ meson with the mass mentioned 
above but also reproduce the phenomenological 
parameters $L_i$ and $H_i$ appearing
in the nonlinear chiral lagrangian \cite{GL} 
up to $O(p^4)$ only to which 
calculations are available\cite{ruiz}; see also 
an excellent review\cite{bijnens}.

It means that the fact that the renormalizable 
linear sigma model
may not match the low-energy phenomenology, as emphasized by 
Gasser and Leutwyler \cite{GL}, do not necessarily deny
the linear realization of chiral symmetry as given in the 
NJL-like models
\footnote{It seems that  the vector terms must be
 incorporated for a complete description of low-energy phenomena
in the NJL model approach \cite{Takizawa,vogl,bijnens}.
}.

\subsection{Chiral symmetry, analyticity, crossing symmetry
 and the $\sigma$-meson pole in the $\pi$-$\pi$ scattering matrix}
 
A tricky point on the $\sigma$ meson is that 
the elusive meson strongly couples 
to two pions to acquire  a large width 
$\Gamma \sim m_{\sigma}$, which makes tough to deduce a
phase shift reliably enough 
of the \pipi scattering in the $I=J=0$ channel.
Nevertheless 
recent cautious phase shift analyses of the \pipi scattering in 
the scalar channel have come to 
claim a $\sigma$ pole of  the scattering matrix 
in the complex energy plane  
with the real part Re\, $m_{\sigma}= 500$-800 MeV and the imaginary 
part Im\, $m_{\sigma}\simeq 500$MeV\cite{pipi}.

One should mention here that 
the same phase shift can be well 
reproduced without the $\sigma$ pole but only with the
$\rho$ meson pole in the $t$-channel using a model 
lagrangian\cite{juerich}.
Then one may naturally wonder whether we can go or not with the $\sigma$.
A nice resolution of this dilemma has been provided by 
Igi and Hikasa\cite{IH};see also the contribution 
by Oller\cite{oller}.

Igi and Hikasa constructed the invariant amplitude for the 
\pipi scattering so that it satisfies the chiral symmetry low energy 
theorem, analyticity, unitarity and approximate crossing symmetry
using the $N/D$ method\cite{ND}.
In construction, they assumed
 possible existence of resonances in $I=J=0$ as well 
as $I=J=1$ channels.
They calculated the two cases with and without the scalar 
resonance by putting on and off the coupling $g_{\sigma}$ of the
$\sigma$ with the pions.
When including the $\sigma$, they assumed that the scalar meson is 
degenerated with the $\rho$ meson as given in the mended symmetry
\cite{weinberg}, which 
also gives the same 
couplings $g_{\sigma}=g_{\rho}$ of the mesons with the pions.
Then their amplitude has essentially  no free parameter;
actually, the coupling constant $g^2_{\rho}$ is slightly 
varied between the
 KSRF coupling $m_{\rho}^2/2f_{\pi}^2$ and 
$m_{\rho}^2/\pi f_{\pi}^2$ given by 
the Veneziano amplitude.
What they found is that the $\rho$ only scenario can account only 
about half of the observed phase shift, while the degenerate
$\rho$-$\sigma$ scenario gives a reasonable agreement with the
data.

In the approach in Ref.17,
 it is unfortunately quite unclear how well 
 chiral symmetry,  analyticity, nor crossing symmetry
 even in an approximate way are taken into account.

\section{Hadron phenomenology and the $\sigma$ meson} 

If the $\sigma$ meson with a low mass is identified, many 
 experimental facts  which otherwise are mysterious 
 can be nicely accounted 
for  in a simple way\cite{HK94,ptp95}.

\subsection{$\Delta I= 1/2$ rule and the $\sigma$ meson}

The correlation in the scalar channel as summarized 
by such a scalar meson may account 
for the enhancement of the $\Delta I=1/2$ processes in 
 K$^{0} \rightarrow \pi^{+}\pi^{-}$ or 
 $\pi^{0}\pi^{0}$ \cite{morozumi}.
In fact,  the final state interaction for the emitted two pions
may include the $\sigma$ pole, then
 the matrix element for of the scalar 
operator $Q_6\sim \bar{q}_Rq_L\bar{q}_Lq_R$ is sown to be
\beq
\la \pi^{+}\pi^{-}\vert Q_6\vert K^0\ra&=&
\la \pi^{0}\pi^{0}\vert Q_6\vert K^0\ra , \\
  \  &=& Y_0\times \frac{F_K}{3F_{\pi}-2F_K}.
\eeq
where the $Y_0$ is the standard matrix element given 
by the vacuum saturation approximation and 
the last factor involving the pion and the kaon decay 
constants gives the enhancement factor due to the $\sigma$ pole.
The relevance of the $\sigma$ pole is best seen by 
rewriting it in  terms the meson masses\cite{morozumi},
\beq
\frac{F_K}{3F_{\pi}-2F_K}=\frac{m_{\sigma}^2-m_{\pi}^2}
                          {m_{\sigma}^2-m_{K}^2}
                          \frac{F_{K}^2}{F_{\pi}^2},
\eeq
which shows that  the approximate degeneracy of the kaon and the $\sigma$
can  give a large enhancement as required to account the 
experimental data.\footnote{
One should, however, notice that  it is urgent
 to explore whether the $\sigma$ can 
systematically describe the processes involving the kaon 
like the mass difference of K$_L$ and K$_S$\cite{terasaki}.
We remark that the
 incorporation of the vector mesons may be a missing link
for the systematic description of the weak processes involving the 
kaon, as in the description of the \pipi scattering in a consistent
way with chiral symmetry, analyticity and crossing symmetry.}

\subsection{The nuclear force and the $\sigma$ meson}

The phase shift analyses of the nucleon-nucleon scattering
 in the $\ ^1S_0$ channel show the existence of the
 state-independent attraction in the 
intermediate range, $1\sim 2$ fm.
This attraction is indispensable for the binding of a nucleus.
In the meson-theoretical models for the nuclear
 force, i.e.,
One-Boson-Exchange Potential (OBEP),
  a scalar and isoscalar meson exchange with the mass range  
400$\sim $ 700 MeV is responsible for
the state-independent attraction\cite{sawada}.
The boson responsible for the state-independent attraction 
has been denoted as ``$\sigma$'' with a quotation, because 
it is a substitute of the two-pion exchange potential in this
 channel; the two-pion exchange includes the {\em ladder},  
the {\em cross} and the {\em rescattering} diagrams
with the $\Delta (1232)$ being incorporated in the intermediate
states.

However, the problem is again how \chis is taken into account 
to construct the N-$\bar{\rm N}$ to \pipi amplitude in the
 $t$-channel.\footnote{A discussion to take into account \chis
 in the nuclear force was first given by Brown\cite{brown}.}
An analysis which respects \chis showed that the direct 
$\sigma$-N coupling is necessary to insure \chis in
the N-$\bar{\rm N}$ to \pipi amplitude\cite{DURSO}.
Furthermore, the amplitude of the 
{\em rescattering} of the pions  mentioned above
should be constructed consistently with the 
\pipi phase shift in the $I=J=0$ channel, for which 
we have seen that \chis , analyticity and the crossing symmetry
are important.\footnote{
In a phenomenological point of view,
 the nuclear forces with and without the direct $\sigma$-N 
coupling are indistinguishable.  Nevertheless
when one tries to describe the 
baryon-baryon interactions including hyperon-N
and hyperon-hyperon interaction systematically,
one may encounter cases where 
the baryon-baryon forces constructed in a chirally symmetric way 
with the direct $\sigma$ baryon coupling give a better 
phenomenology than the ones constructed without the direct
$\sigma$-baryon coupling\cite{tamagaki}.}

\subsection{The $\pi$-N sigma term and the $\sigma$ meson}

The collective excitation in the scalar channel
 as described as   the $\sigma$ meson is essential 
in reproducing the empirical value of the $\pi$-N 
sigma term \cite{JK}
 $\Sigpi =\hat {m}\la \bar{u} u + \bar {d} d\ra$\cite{kuni90}.

The basic quantities here are the 
quark contents of baryons $\la B\vert \bar {q}_i q_i\vert B\ra \equiv
 \la \bar{q}_iq_i\ra_B$ ($i= u, d, s, ...$).  Actually, it is more 
adequate to call them the scalar charge of the hadron. 
Feyman-Hellman theorem tells us that
\beq
 \la \bar{q}_iq_i\ra _B= \frac{\partial M_B}{\partial m_i},
\eeq
which shows that once the baryon mass $M_B$ is 
known as a function of the current quark masses $m_i$, the quark 
content of the baryon is calculable.
The problem is of course to give the functional dependence
 of $M_B$ on 
 $m_i$. 
For this purpose, the chiral quark model \cite{CHQM} is 
useful\cite{kuni90},
 where $M_B$ is given with  ``constituent quark masses'' $M_{i}$ 
 which is identified with 
 the mass  generated by the dynamical breaking of 
 chiral symmetry (plus current quark mass), hence \chis
and the constituent quark model is nicely reconciled in the chiral
quark model. 

One will immediately
 find that  the scalar charge of a baryon 
 is given in terms of the scalar charge of the constituent quark,
\beq
Q_{ji}\equiv \frac{\partial M_{j}}{\partial m_i}=
\la \bar{q}_iq_i\ra _{q_j},\quad
i, j = u,d, s,....  
\eeq
Notice that  
\beq
\frac {d\la \bar {q}_i q_i \ra}{dM_i}= \Pi ^{S}_{i}(q^2=0),
\eeq
where $\Pi ^{S}_{i}(q^2)$ is the zero-th order polarization in the 
scalar channel due to the i-quark (i$=$u, d, s).
If one uses the NJL model with a determinantal 
interaction\cite{KOMAS},
one has \cite{kuni90}
\beq
{\bf Q}=\bigl[{\bf 1}+ {\bf V}^{\sigma }\cdot {\bf \Pi }^{S}(0)
\bigr]^{-1},
\eeq
where
${\bf V}^{\sigma}$ is  the  vertex of the interaction
Lagrangian in the scalar channel 
in the flavor basis;${\cal L}^{\sigma }_{res}=$
$\sum _{i,j=u,d,s}:\bar{q}_iq_i V_{ij}^{\sigma}\bar{q}_j q_ j:$.
The scalar charge matrix is nicely rewritten in terms of the
the propagator ${\bf D}_{\sigma}(q^2)$ of the scalar mesons
\beq
{\bf Q} = - {\bf D}_{\sigma}(0)\cdot {\bf V}_{\sigma}^{-1},
\eeq
where
\beq
{\bf D}_{\sigma}(q^2)= -[{\bf 1} + {\bf V}_{\sigma}\cdot {\bf \Pi}^{S}(q^2)]
^{-1} \cdot  {\bf V}_{\sigma}.
\eeq
Notice that  when the interaction is absent
${\bf Q}= {\bf 1}$.

Effective charges are usually enhanced (suppressed) due to 
collective excitations generated by the attractive (repulsive)
forces.  In the present case, we have of course an enhancement.
The enhancement is caused by the polarization of the vacuum in the
 $I=J=0$ channel; one may use the word ``quantum fluctuation'' 
 for the physical origin of the enhancement. 
The quantum fluctuation is nicely 
summarized  by the scalar meson, i.e., the $\sigma$ meson.

The resulting scalar charges of the proton were 
calculated to be\cite{kuni90}
\beq
 \langle \bar uu\rangle_P = 4.97 \, (2), \quad \ 
  \langle \bar dd \rangle_P = 4.00 \, (1), \quad \ 
 \langle \bar{s}s \rangle_P = .53 \, (0),
\eeq
where the numbers with a paranthesis are the scalar charges 
given by the naive quark model. One can see the collective
effect corresponding to the $\sigma$ meson 
enhances  the quark contents
greatly. It is also to be noted that 
 the strangeness in proton are related
 with the flavor-mixing property of the scalar mesons.
Accordingly, we have
\beq
\Sigpi = 49 {\rm MeV},
\eeq
which is in good agreement with the ``experimental value''\cite{JK}.

\subsection{Remarks}

 We remark also the convergence radius 
of the chiral perturbation theory \cite{CH}  is
 linked with the mass of the scalar meson\cite{HK3e,ptp95}.  
 
The above  facts indicate that the scalar-scalar
correlation is important in the hadron dynamics. This is in a sense
natural because the dynamics which is responsible for the correlations
in the scalar channel is nothing but the one which drives the chiral
symmetry breaking.\cite{ptp95}  

\setcounter{equation}{0}
\renewcommand{\theequation}{\arabic{section}.\arabic{equation}}
\section{Partial chiral restoration and the 
$\sigma$ meson in  hadronic matter}

Although the recent phase shift analyses \cite{pipi} 
of the \pipi scattering 
and the identification of the pole in the $I=J=0$ channel which might be
identified with our $\sigma$ meson are great achievement, 
one must say that it is still obscure whether
the pole really corresponds to the quantum fluctuation of the 
chiral order parameter, i.e., our $\sigma$.
As was  first shown by us\cite{HK85}, 
the $\sigma$ meson decreases the mass (softening) 
in association with the chiral restoration in the hot and/or 
dense medium, and  the width of the meson is also expected to decrease
because the pion hardly changes the mass as long as the system is in 
the Nambu-Goldstone phase. 
Thus one can  expect a  chance to see the $\sigma$ meson as a sharp 
resonance at high temperature and/or density.
  
Some years ago, the present author
proposed several nuclear experiments including
 one using electro-magnetic probes 
to produce the $\sigma$ meson in nuclei, thereby 
have a clearer evidence of  the  existence of 
the $\sigma$ meson and also explore the possible 
restoration of chiral  symmetry in the nuclear
 medium\cite{ptp95,tit}.
To make a veto for the two pions from the rho meson, the produced
pions should be neutral ones which may be detected through
four $\gamma$'s.

When a hadrons is put in a nucleus,
the hadron may dissociate into complicated
excitation to loose its identity in the  medium. 
Then the most informative quantity is the response function 
or spectral function of the system.
A response function in the  energy-momentum
space is essentially the spectral function in the meson channel.
If the coupling of the hadron  with the environment is relatively small,
then there may remain a peak with a small width in the spectral function, 
corresponding to the hadron.
Such a peak is to be identified with
 an elementary excitation or a quasi particle, known in 
Landau's Fermi liquid theory for fermions. 
It is quite nontrivial whether a many-body system can
admit an elementary excitation or quasi-particle
with a specific quantum number. 
Landau gave an argument that 
there will be a chance to describe a system as 
an assembly of almost 
free quasi-particles owing to the Pauli principle 
when the temperature is low\cite{landau}. 
Then how will the decrease of $m_{\sigma}$ in the nuclear medium
\cite{HK85}
reflect in the spectral function in the $\sigma$ channel?

It has been shown by using linear sigma models 
that an enhancement in the spectral function in the
 $\sigma$ channel occurs just above the
two-pion threshold  along 
with the decrease of $m_{\sigma}$\cite{CH98}.
Recently, it has been shown \cite{HKS} that the spectral enhancement
near the $2m_{\pi}$ threshold takes place 
in association with  partial restoration of \chis  
at finite baryon density.

Referring to \cite{HKS} for the detailed account,
we here describe the general
features of the spectral enhancement near the two-pion threshold.
Consider the propagator 
 of the $\sigma$-meson at rest in the medium :
$D^{-1}_{\sigma} (\omega)= \omega^2 - m_{\sigma}^2 - $
$\Sigma_{\sigma}(\omega;\rho)$,
where $m_{\sigma}$ is the mass of $\sigma$ in the tree-level, and
$\Sigma_{\sigma}(\omega;\rho)$ is 
the loop corrections
in the vacuum as well as in the medium.
 The corresponding spectral function is given by 
\beq
\rho_{\sigma}(\omega) = - \pi^{-1} {\rm Im} D_{\sigma}(\omega).
\eeq
One can show that 
${\rm Im} \Sigma_{\sigma}$
$\propto \theta(\omega - 2 m_{\pi})$
$ \sqrt{1 - {4m_{\pi}^2 \over \omega^2}}$
near the two-pion threshold  in the one-loop order.
On the other hand, partial restoration of \chis 
implies that $m_{\sigma}^*$ 
 defined by
 ${\rm Re}D_{\sigma}^{-1}(\omega = m_{\sigma}^*)=0$
  approaches to $ m_{\pi}$.  Therefore,
 there exists a density $\rho_c$ at which 
 ${\rm Re} D_{\sigma}^{-1}(\omega = 2m_{\pi})$
 vanishes even before the complete restoration
 of \chis where $\sigma$-$\pi$
 degeneracy is realized.
At this point, the spectral function is solely 
given in terms of the
 imaginary part of the self-energy;
\beq
\rho_{\sigma} (\omega \simeq  2 m_{\pi}) 
 =  - {1 \over \pi \ {\rm Im}\Sigma_{\sigma} }
 \propto {\theta(\omega - 2 m_{\pi}) 
 \over \sqrt{1-{4m_{\pi}^2 \over \omega^2}}},
\eeq
which cleary shows the near-threshold enhancement of
the spectral function.
 This is a general phenomenon correlated with the 
 partial restoration of \chis.

This result is interesting in relation with
the experiment by CHAOS collaboration  \cite{chaos};
see \cite{HKS,wambach} and the report presented by
 Hatsuda\cite{hatsuda} for more details and recent development.

\section{Summary}

The $\sigma$ meson
is the quantum fluctuation of the amplitude of the 
order parameter of the chiral transition in QCD.
The existence of the $\sigma$ meson does not
 contradict with the success of the chiral perturbation
theory for the low-energy phenomena.
If analyticity, crossing symmetry are 
respected as well as chiral symmetry (\chis) and unitarity,
the phase shift analyses of the \pipi scattering 
in $I=J=0$ channel is  in favor of the existence of 
a scalar-isoscalar meson as well as the $\rho$ meson in $t$-channel.
The scalar meson might be identified with the $\sigma$ meson as
the quantum fluctuation of the chiral order parameter, though
some work is still needed to
make the identification conclusive.
 
If the $\sigma$ meson exists, 
the collective mode in the $I=J=0$ channel as summarized by 
the $\sigma$ can account for various phenomena 
in hadron physics which otherwise
remain mysterious.

The study of the spectral function in the $\sigma$ channel
obtained for the systems with finite  $T$ and/or the 
density $\rho_B$ is interesting 
to elucidate the existence of the $\sigma$ meson more
clearly and also the possible partial restoration of \chis.
The  CHAOS\cite{chaos} group may have seen an evidence of
partial restoration of \chis in the nuclear medium.

\vspace{0.3cm}
This work is supported by the Grants-in-Aids of the Japanese
Ministry of Education, Science and Culture (No. 12640263 and 
12640296).

\end{document}